# Combinatorial Approach to Object Analysis

November 4, 2005

**Rami Kanhouche**[1]

## Abstract

We present a perceptional mathematical model for image and signal analysis. A resemblance measure is defined, and submitted to an innovating combinatorial optimization algorithm. Numerical Simulations are also presented.

.

## Introduction.

Object Analysis, from this paper point of view, is just a continuity to the already well defined Object Oriented Programming and modeling techniques, with a difference, that is, we will be looking for automated methods realizing the analysis of the object, and eventually construct an object model of a given environment –or a signal. From one hand the "Object" concept define a central point for Object's Data storage, and the functions, interfacing it to the external world, and on the other hand, the "Object" concept, threw its hierarchy, is an actual investment of "similarities" between different object forms, known as *polymorphisms*. Object programming has been used, with a great success, in computer science. But the thinking process, or the analysis process, generating these models, is of course nothing but intelligence; our intelligence, with its inherent complexity. In our search for an automated object-analysis capable algorithms –or machines, image processing, and more generally signal processing, are the most capable in what we know in science. To this date, image-processing science, coupled to the information processing science, do provide us with different analysis technique of the signal that can be categorized into these categories:

1. **Low Level processing**: dealing with signal directly, with no consideration of the Object characteristic of the signal, i.e.: signal compression, signal enhancement.
2. **High Level processing**: dealing with the signal after some quantification process, i.e: tresh-holding. In this category of methods, the object concept is rarely taken into consideration, and the classification techniques used for learning "New Objects" into the system are totally strange to our natural way of Object perception. Also, the quantification process results in a great loss of information.
3. **"in between" methods**: i.e. Neural Network, Splines, these methods are used in reproducing an intelligent behavior, threw intensive learning approach. The essential point of these methods is the investment of the non-linear mathematics as a main approach to reproducing an intelligent activity.

---

[1] PhD Student at Lab. CMLA, Ecole Normale Supérieure de Cachan, 61, avenue du Président Wilson, 94235 CACHAN Cedex, France. phone: +33-1-40112688, mobile: +33-6-62298219, fax: +33-1-47405901, e-mail: rami.kanhouche@cmla.ens-cachan.fr, kanram@free.fr.

Methods in this category were so successful in simulating different intelligent application, but suffered also from the following disadvantage:

- Incapability to generalization: the quality of the output is inversely proportional to the distance between the "new input" and the already learned "inputs".
- As in category 1's methods, there is no conscience about the Object concept.

In this paper, we will not explain how to make an intelligent process, or give a magical algorithms, but instead we will explain the philosophical principles that is, to our eyes, essential to realize such thing. And building on that, we will propose a new mathematical approach that confirms our point of view.

## I. Fundamental Abstraction Principle.

*"Abstraction –or resemblance- is relative."*

Contrary to what we used to do until now, any object resemble to other objects not by a "number" that is quantifying its degree of resemblance, instead any two objects do resemble according to all possible sub-objects resemblances, and this is a space of extreme possibilities, and only the investment of these sub-spaces of sub-objects resemblances can be able to give an accurate perceptional measure of the original two object's resemblance. This investment of the sub-objects is what we can call "combinatorial perception". And is formalized into the following:
Any given signal $I_1(x): R^n \to [0,1], x \in R^n$, resemble to a second signal $I_2(x): R^n \to [0,1]$, with a different degrees, proportional to the minimums of the function

$$\aleph(I_1, I_2) := \frac{1}{|\Im_1|} \sum_{x \in \Im_1, y \in \Im_2} \left| \frac{dist(x,y)}{\max\{dist(x,y)\}} - \frac{dist(\Omega(x),\Omega(y))}{\max\{dist(\Omega(x),\Omega(y))\}} \right| \qquad (1),$$

subject to $\Omega : \Im_1 \to \Im_2$, where $\Im_u := \{x : x \in R^n, I_u(x) \neq 0\}$ .

In the previous problem description we considered any signal of the nature $R^n \to R$, as a hyper-shape, or form, inside the space $R^{n+1}$. The $dist(x,y)$, can be taken as the simple Cartesian distance, i.e. when $x, y \in R^n$, $dist(x,y) := \left( \sum_{i=0}^{n-1} [x_i - y_i]^2 \right)^{\frac{1}{2}}$. In the case when the transformation $\Omega$ is a bijection between two images of the same size, we can formulate the problem in matrix notations as

$$\min_W \sum_{i,j} \left\lfloor \left| D_2 - W^T D_1 W \right| \right\rfloor_{ij} \qquad (2),$$

where $D_1, D_2$ are the auto-distance matrices for the signal $I_1$, and $I_2$ ,respectively. By taking a space- numbering function $f(i) := x_i : N \to \Im_u$, then the auto-distance matrix is defined as $D_u := \lfloor d_{ij} \rfloor_{i,j:=0..|\Im_u|-1}$, with $d_{ij} := dist(x_i, x_j)$. On the other hand the operator $W$, is a matrix of size $|\Im_1|\times|\Im_1|$, defined as

$$W := [w_{ij}]_{i,j:=0\ldots|\Im_1|-1}, w_{ij} = \begin{matrix} 1 & if\ j=\Omega(i) \\ 0 & if\ not \end{matrix}. \qquad (3)$$

According to the proposed principle we will have to deal with a combinatorial space, of permutations with repetition for (1), and without repetition for (2). In this paper we will only treat the situation for permutations without repetition according to model (2). For that we will note the permutation without repetition, of order $k$, from the group $N_1 := \{0,1,\ldots k-1\}$, into the group it self as the bijection $\pi : N_1 \to N_1$. The next proposition is of great practical usefulness for the control over "all permutations" space $\{\pi_i\}_{i=0\ldots k!-1}$. More precisely we will be projecting these permutation on the line.

**Lemma 1.**

$$\forall g > 1, g! > (g-1)!(g-1) + (g-2)!(g-2) + \ldots 1. \qquad (4)$$

*Proof.* By taking the sum $(g-1)!(g-1)$, to the left of the relation, according to $\forall g > 1, [g-(g-1)](g-1)! > (g-2)!(g-2)+\ldots 1$, we find $\forall g > 1, (g-1)! > (g-2)!(g-2) + (g-3)!(g-3) + \ldots 1$, By continuing recursively the relation do realize.

**Proposition 1.** There is a bijection between the space of all permutations without repetition of order $k$ and the line according to

$P(\pi_i) : \{\pi_i\} \to \{0,1\ldots k!-1\}$, where

$$P(\pi_i) := |k-1|!\pi_i'(k-1) + |k-2|!\pi_i'(k-2) + \ldots + 2!\pi_i'(2) + 1.\pi_i'(1) \qquad (5)$$

where $\pi_i'(l) := \pi_i(l) - |s_l|$, and $s_l := \{\pi_i(b) : b > l, \pi_i(b) < \pi_i(l)\}$. (6)

Proof. At first we will prove invertible, the transformation from the space $\{\pi_i\}$, to the more compacted manner of coding the permutations using $\pi_i'$, where according to (6) we got always $\pi_i'(l) \in \{0,\ldots l\}$. It is sufficient to notice, that starting from $\pi_i'(l)$, $l = 0,1\ldots,k-1$, and in the inverse order from $l = k-1$, down to $l = 0$, we can write $\pi_i(l) = \pi_i'(l) + |s_l|$, with the values $\pi_i(b)$, $b > l$, needed to the definition of $s_l$, always available at step $l$.

Next, according to *Lemma 1*, we got always $0 \le P(\pi_i) <= k!-1$, while for the bijection between the space $\{\pi_i'\}$, and $\{P(\pi_i)\}$, we will prove that

$\forall \pi_i', \pi_j', \forall l, \pi_i'(l) \neq \pi_j'(l) \Rightarrow P(\pi_i') \neq P(\pi_j')$.

Supposing that for $\pi_i', \pi_j'$, $\pi_i'(l) = \pi_j'(l)$, for $l > y$, with $\pi_i'(y) \neq \pi_j'(y)$ , then $\forall \pi_i'(w), \pi_j'(w), y > w$,

$P(\pi_i') - P(\pi_j') = (y)!(\pi_i'(y) - \pi_j'(y)) + (y-1)!(\pi_i'(y-1) - \pi_j'(y-1)) + \ldots$, since that $\pi_i'(t) \in [0,1\ldots t]$, we can write

$$\begin{aligned} &|(y-1)!(\pi_i'(y-1) - \pi_j'(y-1)) + (y-2)!(\pi_i'(y-2) - \pi_j'(y-2)) + \ldots| \\ &\le (y-1)!(y-1) + (y-2)!(y-2)\ldots + 1 \end{aligned}.$$

Next, by *Lemma 1*, we arrive to find $\begin{matrix} \left|(y-1)!\left(\pi'_i(y-1)-\pi'_j(y-1)\right)+(y-2)!\left(\pi'_i(y-2)-\pi'_j(y-2)\right)+\ldots\right| \\ < \left|(y)!\left(\pi'_i(y)-\pi'_j(y)\right)\right| \end{matrix}$, from which $P(\pi'_i)-P(\pi'_j)\neq 0$. This also leads us directly to the result $\forall P(\pi'_i), P(\pi'_j),\ P(\pi'_i)=P(\pi'_j)\Rightarrow \forall l, \pi'_i(l)=\pi'_j(l)$. What remains to prove the bijection is that $\forall\ v\in\{0,1,\ldots k!-1\}, \exists \pi' : v=P(\pi')$, the proof comes from the fact that there is exactly $k!$ elements on the line, $\{P(\pi_i)\}$, which are the projections of exactly $k!$ known permutation without repetition, for that there could not be any element not so.
□.

As we already mentioned, the space of possible combinations relative to the optimization of (1), or even (2), is very huge from calculus point of view. In the future, different ways of approaching this space can be proposed. For the moment, in our exploration effort for methods investing our model, we found it most adequate to proceed according to the following:

1- Select a small number, $L$, of equally distributed points of each image.
2- Construct the auto-distance matrix $D_1$, and $D_2$.
3- Instead of scanning the space of all possible $L!$, permutations, we will scan the space on equal interval of size $J$, or what we can call *Jumping*. Proposition 1 is very essential to be capable of realizing such a way of scanning.
4- Starting from the quasi-minimums localized threw step 3, we will proceed into an optimization process, which got the character of having *Minimum Inertia*. We mean by *Minimum Inertia*, that the optimization Algorithm, which we will be explaining in the next Section, will try to find a relatively "local minimum", in the sense of the most minimum changes in the solution $W$ matrix.

## II.Minimum Inertia Relative Resemblance Optimization Algorithm.

**Definition 1**. For any given permutation $\Omega_i : L \to L$, we call the space of permutation changes concerning exactly $O$ elements, as the $\Delta(\Omega_i, O)$ Level Space.

**Definition 2.** For any given permutation $\Omega_i : L \to L$, we call the sub space of circulate permutation changes concerning exactly $O$ elements, as the $\Delta^C(\Omega_i, O)$ Level space.

**Proposition 2.** For any $O' > O$, $\Delta(\Omega_i, O')\setminus\Delta(\Omega_i, O)=\Delta^C(\Omega_i, O')$.

For an example, for the permutation $\Omega_i = \{0\to 1, 1\to 2, 2\to 0\}$, the Level space $\Delta(\Omega_i, 2)$ contains the changes $[0\to 2, 1\to 1]$, $[0\to 0, 2\to 1]$, and $[2\to 2, 1\to 0]$.

The Minimum Inertia Optimization Algorithm that we are proposing, perform as the following:

*" Starting from Level Space 2, test changes of the permutation $\Omega_i$, and apply them if they only minimize (2), when all the Level space is tested with no changes applied, proceed to the higher level."*

When a minimization occurs in a higher level $\Delta(\Omega_i, O)$, and according to our minimum inertia criteria, our choice is to restart from level 2, because that a higher level's change represent a new major orientation in the perceptional correspondence, which in turn, need to be more investigated. Here is the formalization of the proposed algorithm:

*Note $W$, calculated threw $\Omega$, as $W^{\Omega}$.*
*Define* $f(\Omega) := \sum_{i,j}\left[\left|D_2 - \left(W^{\Omega}\right)^T D_1 \left(W^{\Omega}\right)\right|\right]_{ij}$
*Note $\Omega^{\ell}$, as the permutation $\Omega$, after applying the Level changes $\ell \in \Delta(\Omega, O)$.*
*For a starting permutation of size $L$, set the maximum Levels Count as $2 \le H \le L$.*
*Set h=2*
*Start:*
*For all $\ell \in \Delta^C(\Omega_i, h)$ {*
*$f(\Omega_i^{\ell}) < f(\Omega_i)$ {*
*set $\Omega_i = \Omega_i^{\ell}$*
*set $h = 2$*
*goto Start*
*}*
*}*
*if( $h < H$ ) {*
*$h = h + 1$*
*goto Start*
*}*
*else {*
*END.*
*}*

**III. Numerical Simulations.**

In this section we will be looking for the application of what proceeded. While we are confident that the proposed resemblance principle is totally capable of analyzing direct signals, or images, the tested numerical images were selected as binary images. The reasons behind our choice are, from one hand, the enormous calculus cost needed for real signal, and the clarity of the theory explanation provided by using binary images, on the other hand.
In the construction of the solutions images, only a few points of the source image $I_1$, and the target image $I_2$ were selected. Because of that, and to find a generalization of the minimum corresponding to a permutation $\Omega$, over all image points, we proceeded according to the following manner:
For a given permutation $\Omega_i : \Im_1^L \to \Im_2^L$, where $\Im_1^L \in \Im_1$ is the selected group of $L$ points, for each point in the *Source Image* $t_1 \in \Im_1$, we define $\Omega_i' : \{\Im_1^L, t_1\} \to \{\Im_2^L, t_2\}$, with $\Omega'_i(l) = \Omega_i(l)$, for $0 \le l < L$. And select $t_2 \in \Im_2$, between all possible *Target Image* points, which minimizes $f(\Omega'_i)$.
The result of the previous procedure, establishing the connection between the *Source Image* and the *Target image*, are displayed in a *Connection Image*, in which colors were used to express the *Source Image*'s point position information, on the *Target Image*, with green, and blue, for the vertical, and horizontal direction, respectively. Practically it is so probable to have more than a point in the *Source Image* effected to the same *Target Image*'s point, in that case, a mean value is calculated and expressed threw the color component.
To be able to visually evaluate the results presented in the *Connection Image,* a *Colored Source Image* is also presented, in which each bit position coordinates were coded into colors components, by the same way as the *Connection Image*.
Also, in the results that we will be showing, a *Crossing Image*, showing the *Source Image* and the *Target Image*, with curves connecting the $\Im_1^L, \Im_2^L$ groups' points, each marked with a drawn circle.

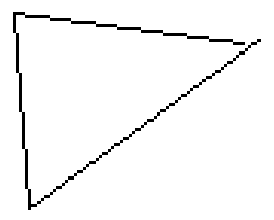

*Figure 1: Triangle (Target)*

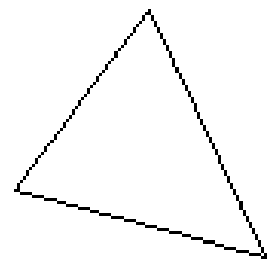

*Figure 2: Triangle (source)*

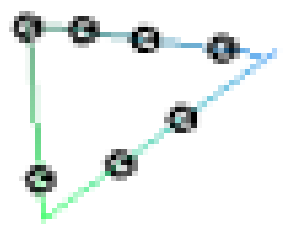

*Figure 3*

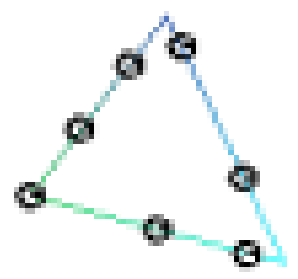

*Figure 4*

In Figure 1, and 2, we see Target and Source Images respectively, for the triangle shape. In Figure 3, and 4 we see a selection of 7 *Root Points*, for each, of the source, and the target. For this simple shape there was no optimization, instead all the possible $7!$ permutations were tested. In Figure 5, the cost function values are displayed, for the permutations according to their projection on the line. The same values were sorted and displayed in Figure 6. We observe the very sharp slope of the curve's start, especially, in the region marked with a drawn box. The sharpness of the slope over a region, denser than the end-region of the curve, is manifestation of we can call a *resemblance region* between the two triangle objects.

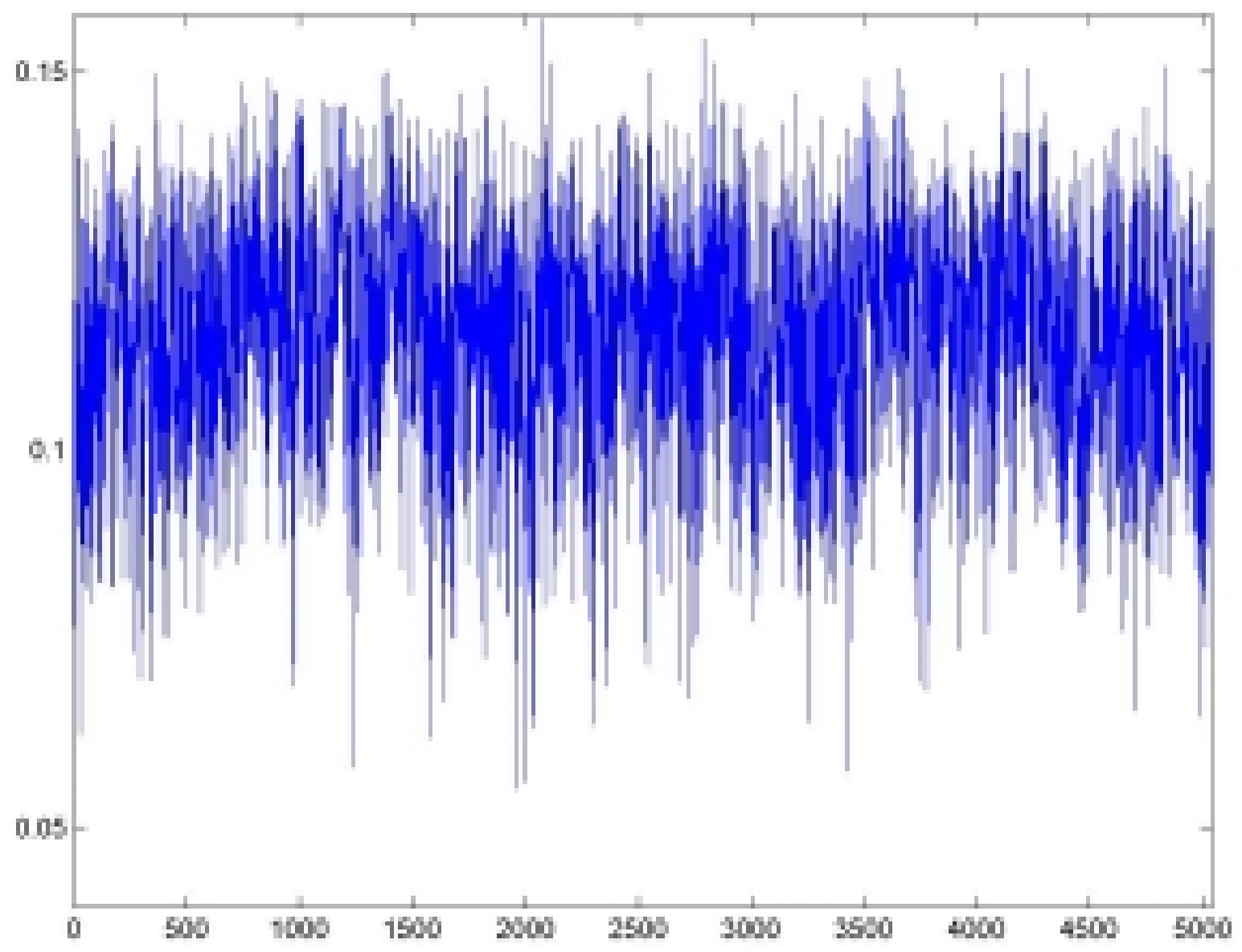

*Figure 5*

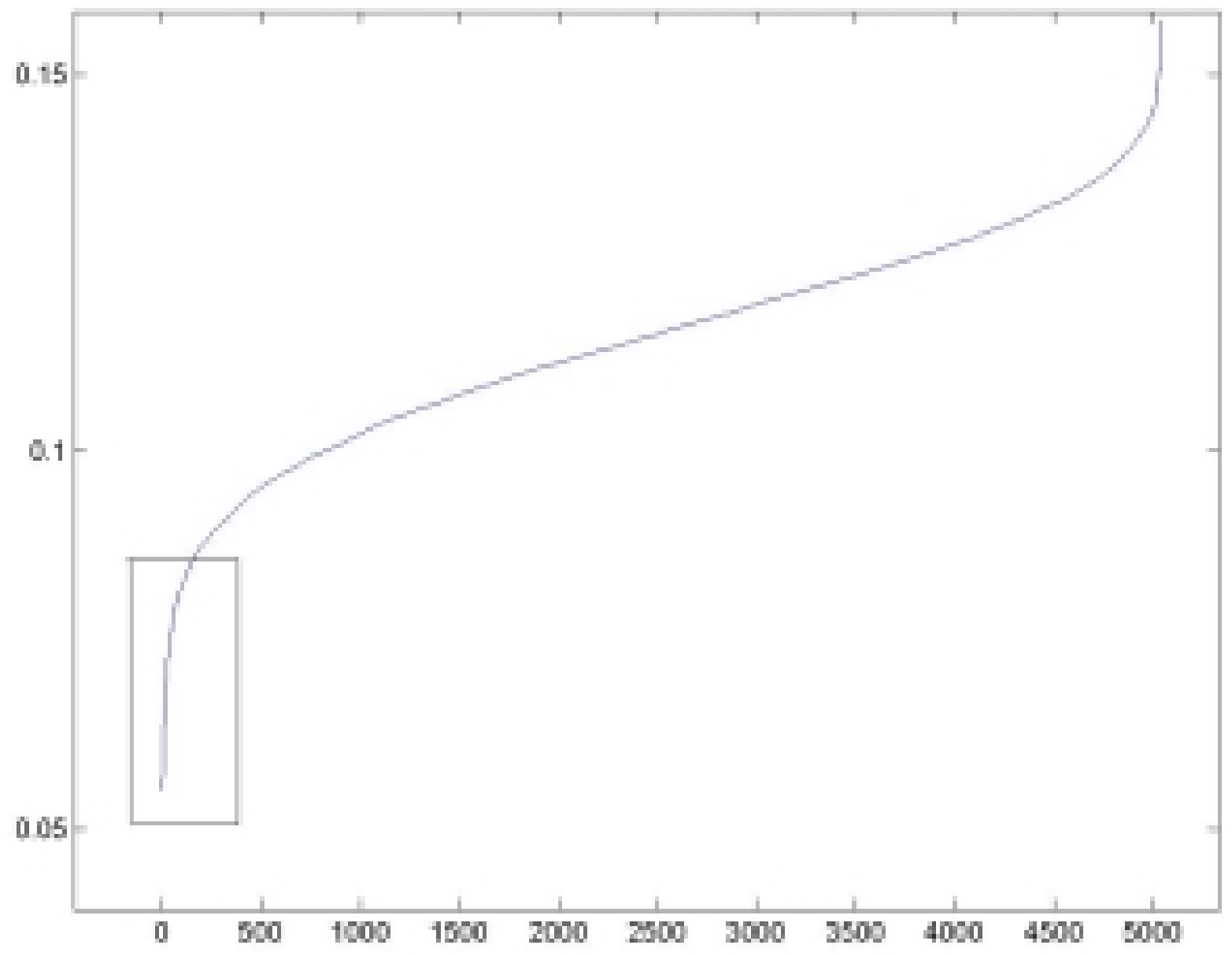

*Figure 6*

As one could expect, there should be 6 minimums in the resemblance search between the two Triangles. By looking in the sorted values solutions, these solutions were found, according to the sorted order, at locations 0, 2, 4, 5, 9,and 21. The Connection Images, and Crossing Images for these solutions are displayed in Figures 7-18.

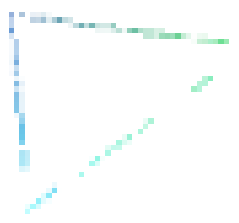

*Figure 7*

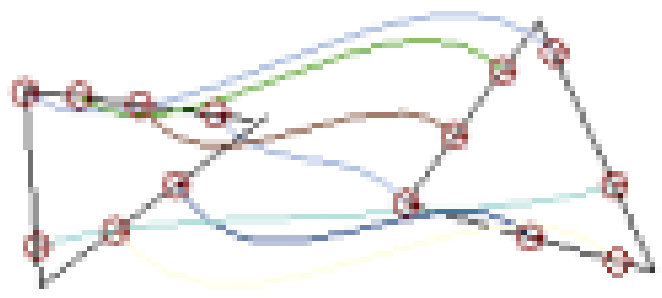

*Figure 8*

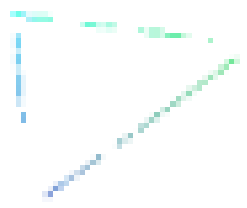

*Figure 9*

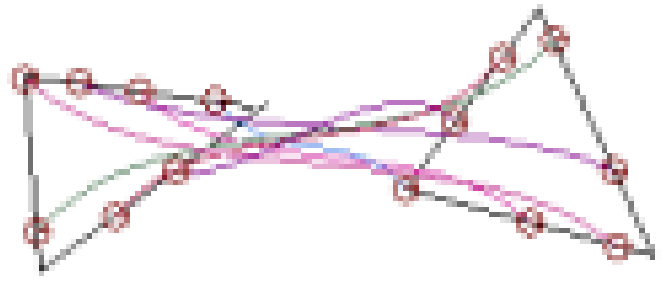

*Figure 10*

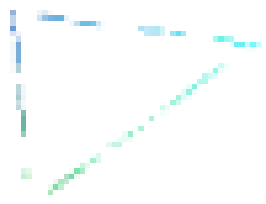

*Figure 11*

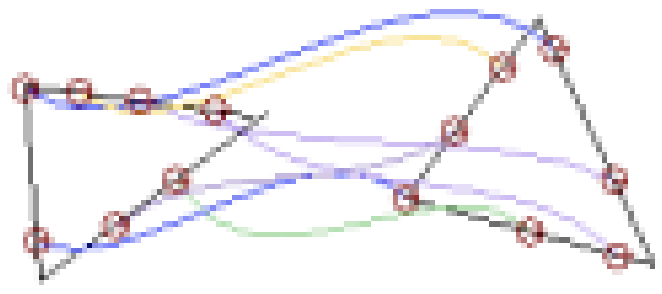

*Figure 12*

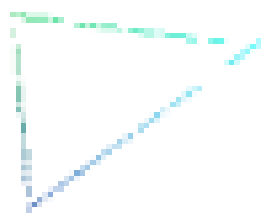

*Figure 13*

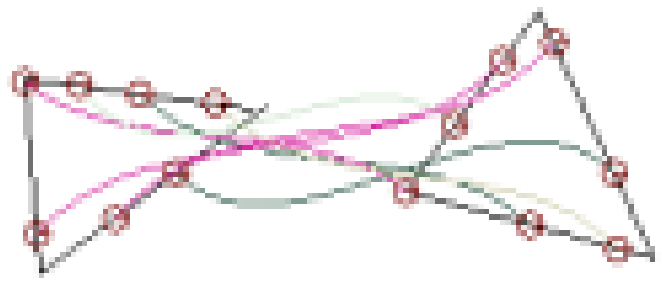

*Figure 14*

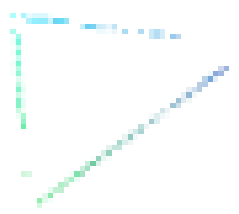

*Figure 15*

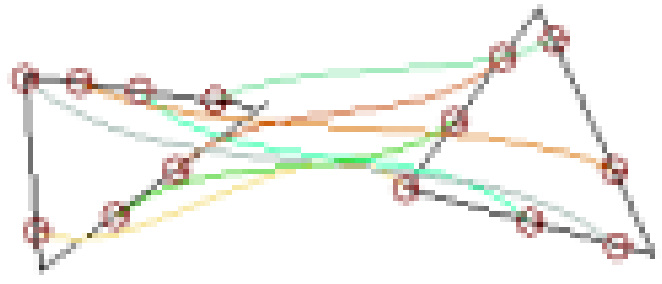

*Figure 16*

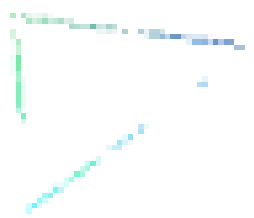

*Figure 17*

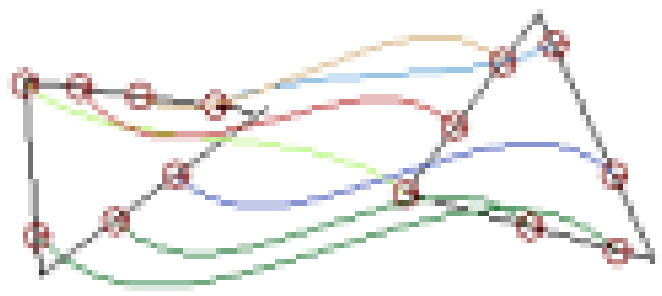

*Figure 18*

Next, and processing more complicated objects; we applied the same procedure on the word form “hello”, with the Target Image, and the Source Image, shown in Figure 19, and 20, respectively.

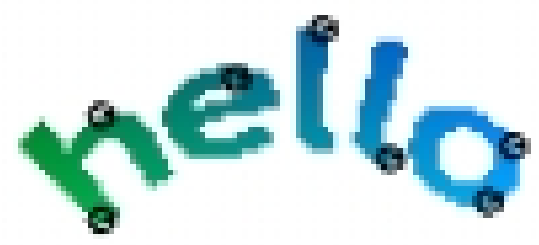

*Figure 19*

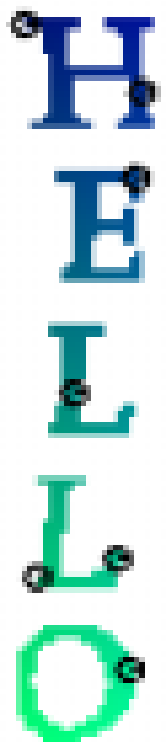

*Figure 20*

Again, looking in the most minimum values, we found the first and second most minimum, with crossing images displayed in Figures 21, and 22.

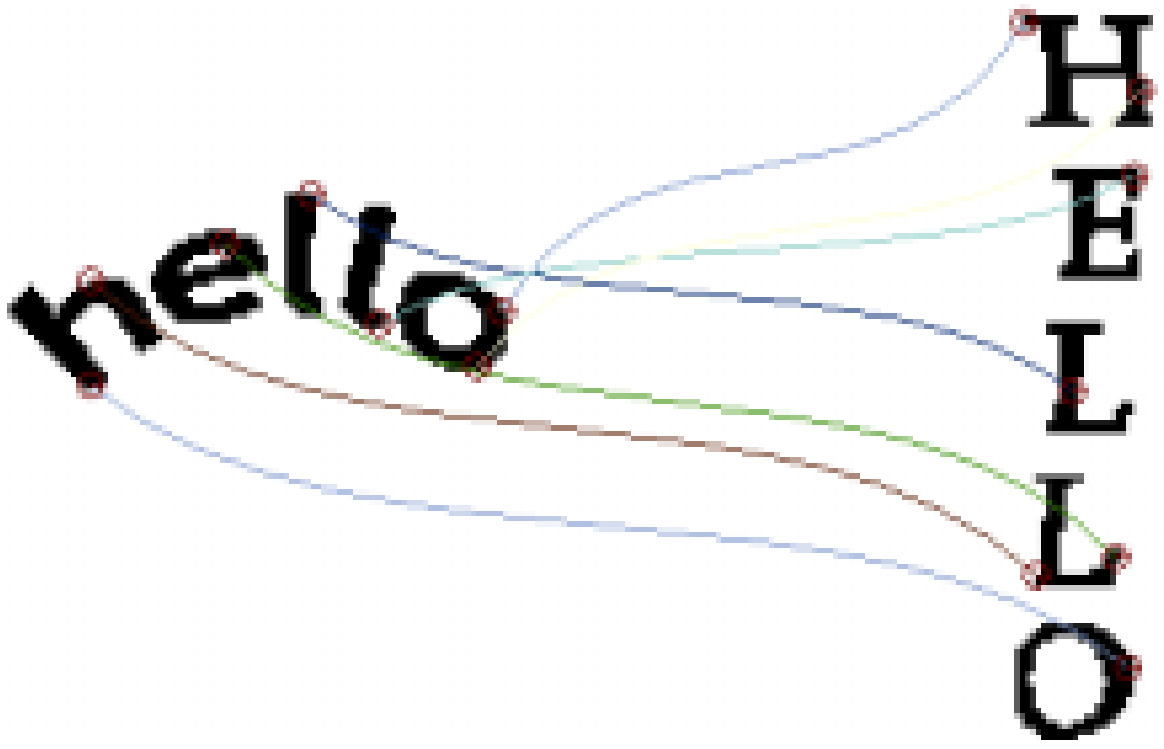

*Figure 21*

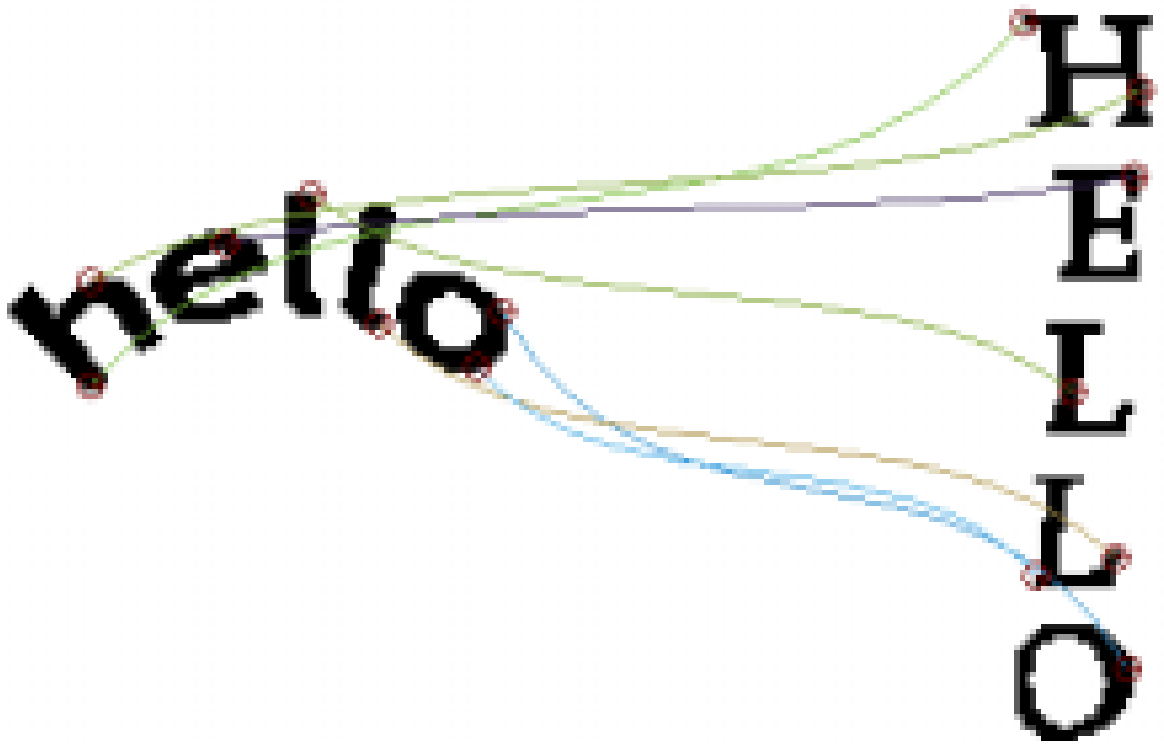

*Figure 22*

At a second step, and starting from the two solutions found using only 7 points, a new number of points were added, to have a total of 15 points in each image, which in turn were submitted to the *Minimum Inertia Optimization algorithm,* using parameter value $H = 10$. The Corresponding Crossing Images before and after optimization are displayed in Figures 23, and 24, for the first solution, and in Figures 25 and 26 for the second solutions.

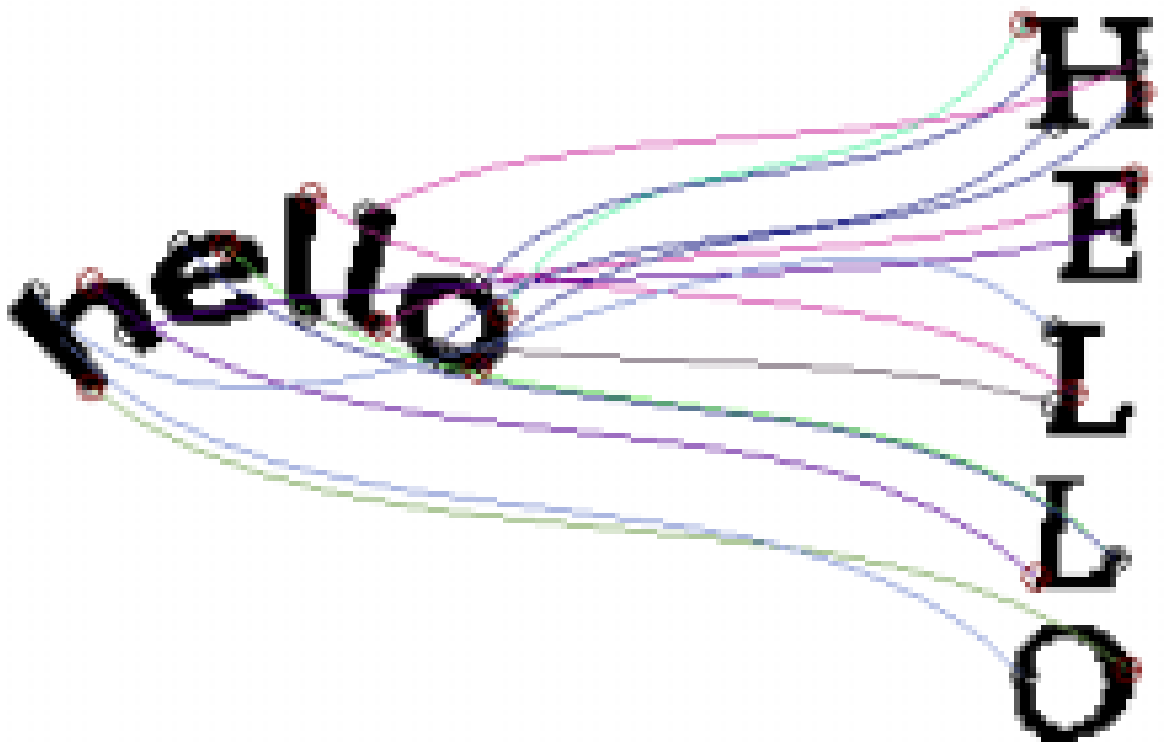

*Figure 23*

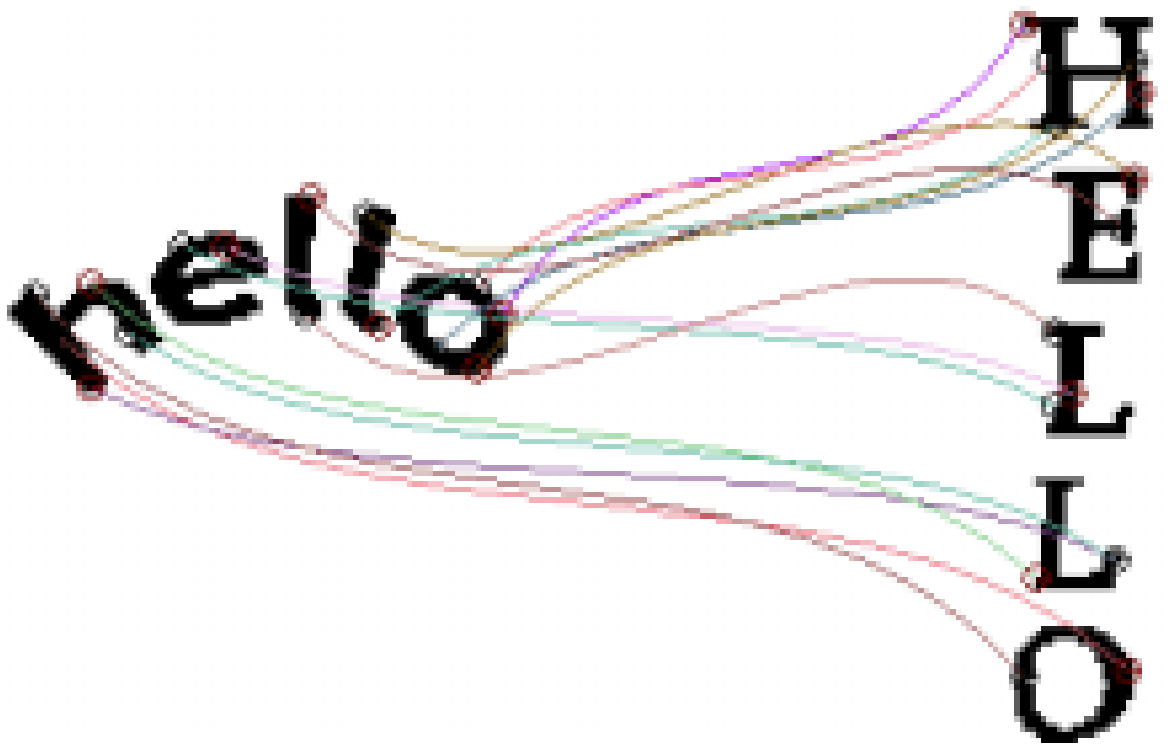

*Figure 24*

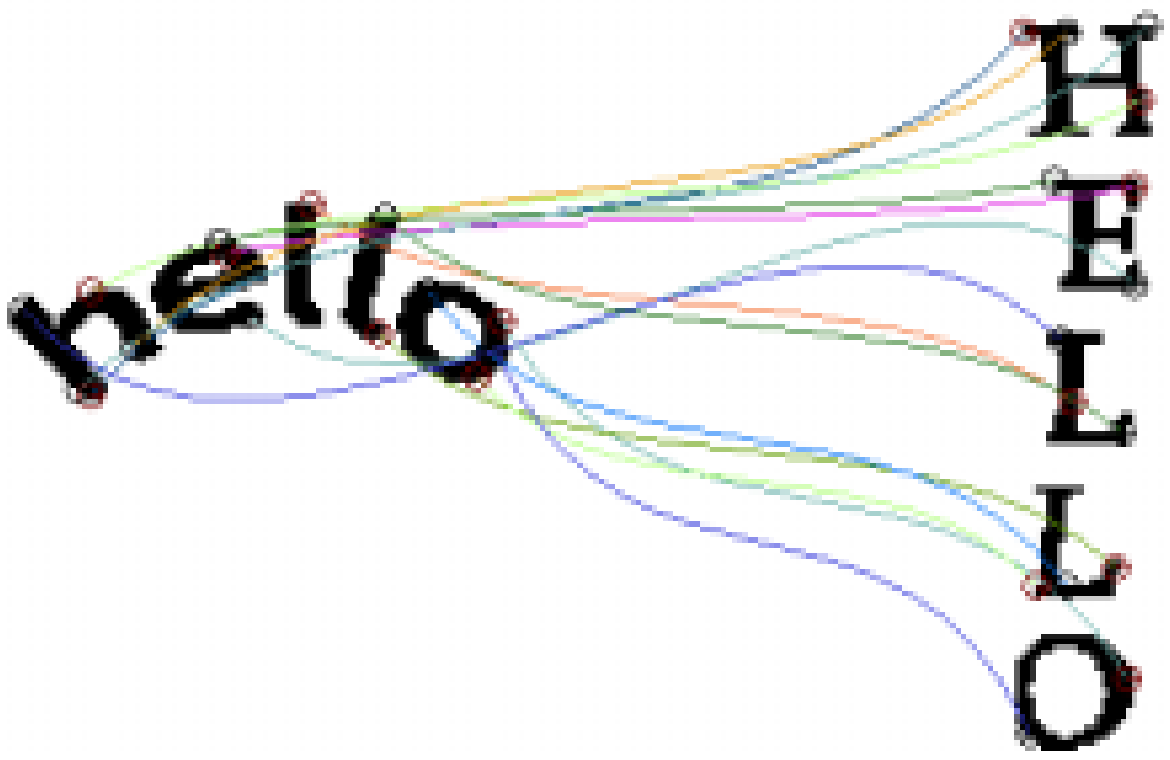

*Figure 25*

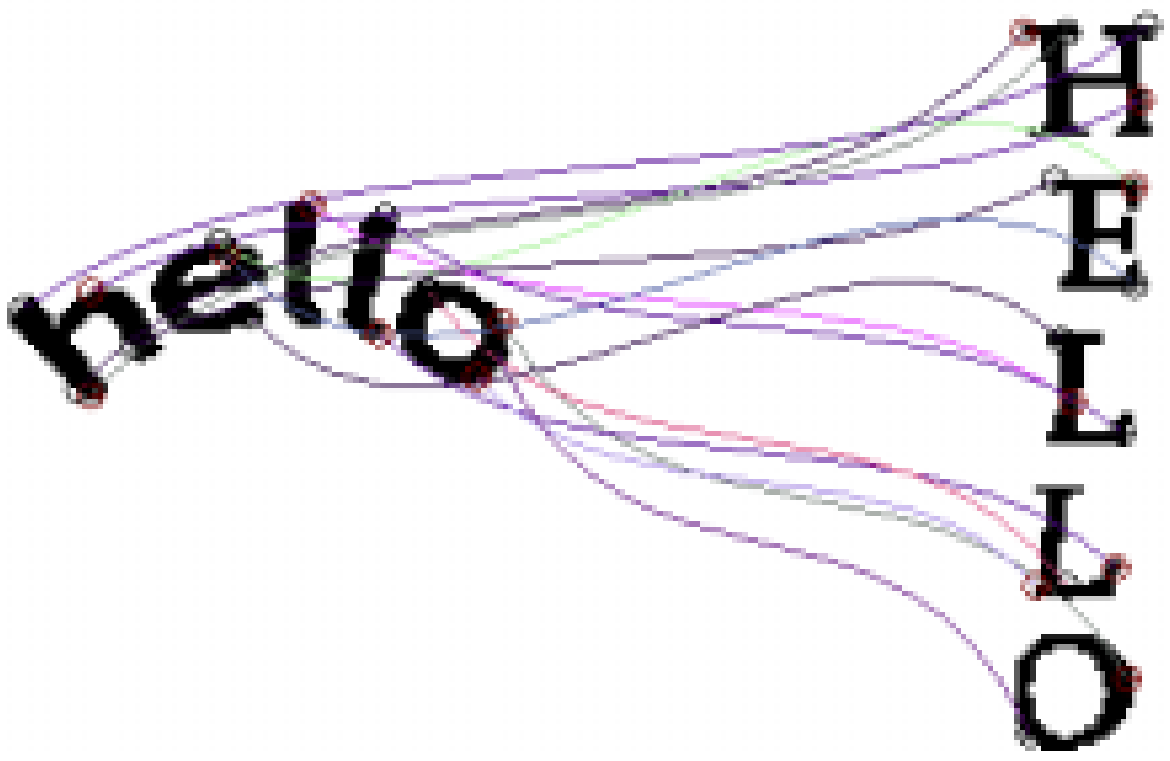

*Figure 26*

Numerically, the result shown in Figure 24 correspond to a cost function value of 0.031, which was optimized starting form the value 0.086, while for the second solution the optimization minimized the value from 0.059 toward 0.037.

| | First Solution | Second Solution |
|---|---|---|
| 7 Points | 0.04719226242 | 0.04760848459 |
| 15 Points (Before Optimization) | 0.08668364031 | 0.0592063447 |
| 15 Points (After Optimization) | 0.0307841795 | 0.03763321309 |

*Table 1*

In *Table I*, we show cost function values for the different steps corresponding to the "hello" example. In Figures 27, and 28, we see the Connection Image and the Crossing Image, respectively, corresponding to the most minimum solution to the Source and Target Images displayed In Figures 29, and 30.

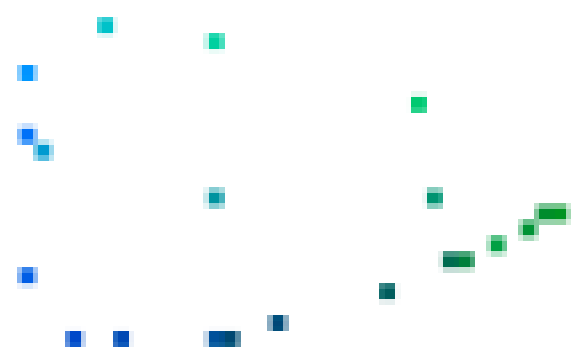

*Figure 27*

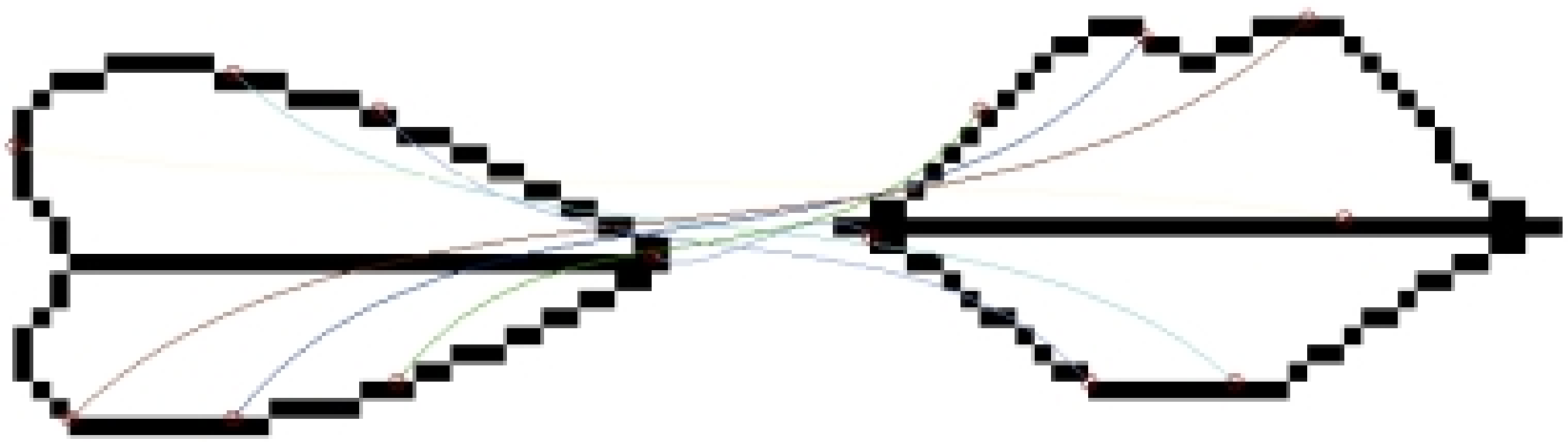

*Figure 28*

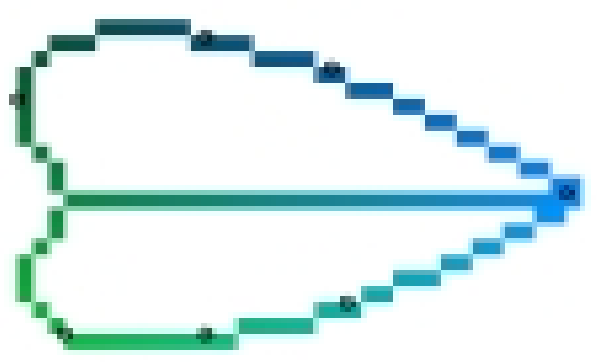

*Figure 29*

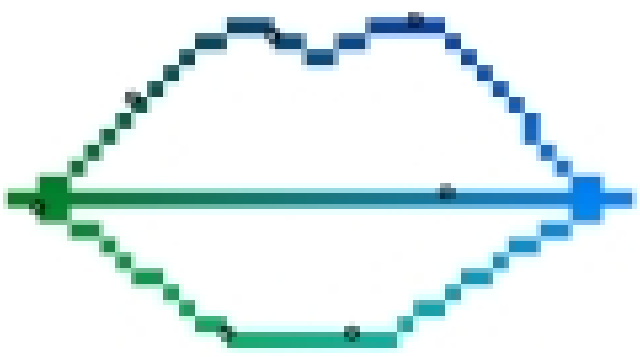

*Figure 30*

The second most minimum 7-point solution to the same example is displayed in the *Figures 31*, and *32*.

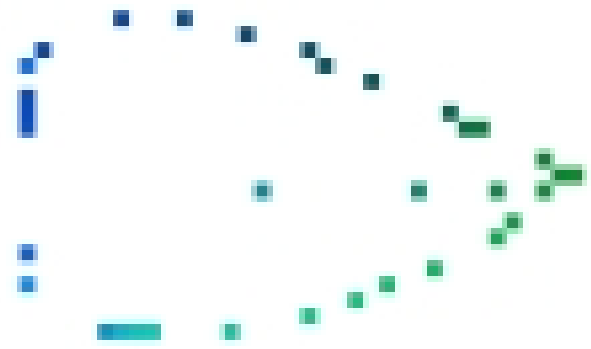

*Figure 31*

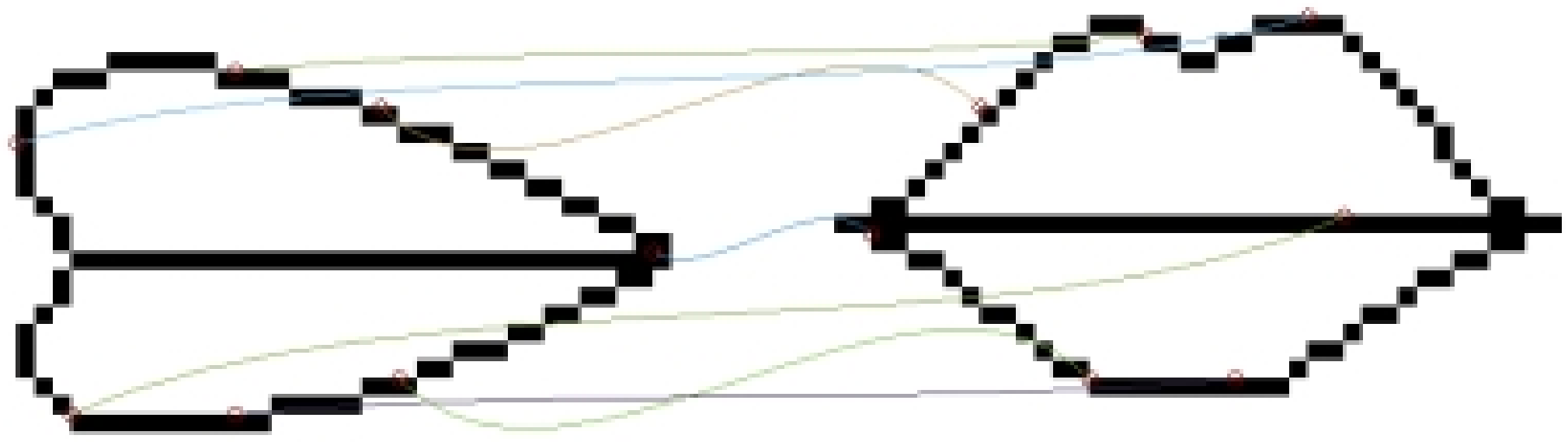

*Figure 32*

As in the previous examples, the Points Number was augmented up to 15 points, and an optimization was initiated for both solutions. Resulting solutions are displayed in Figures 33, 34, and 35, for the First most minimum Solution, and in Figures 36, 37, and 38 for the second most minimum solution.

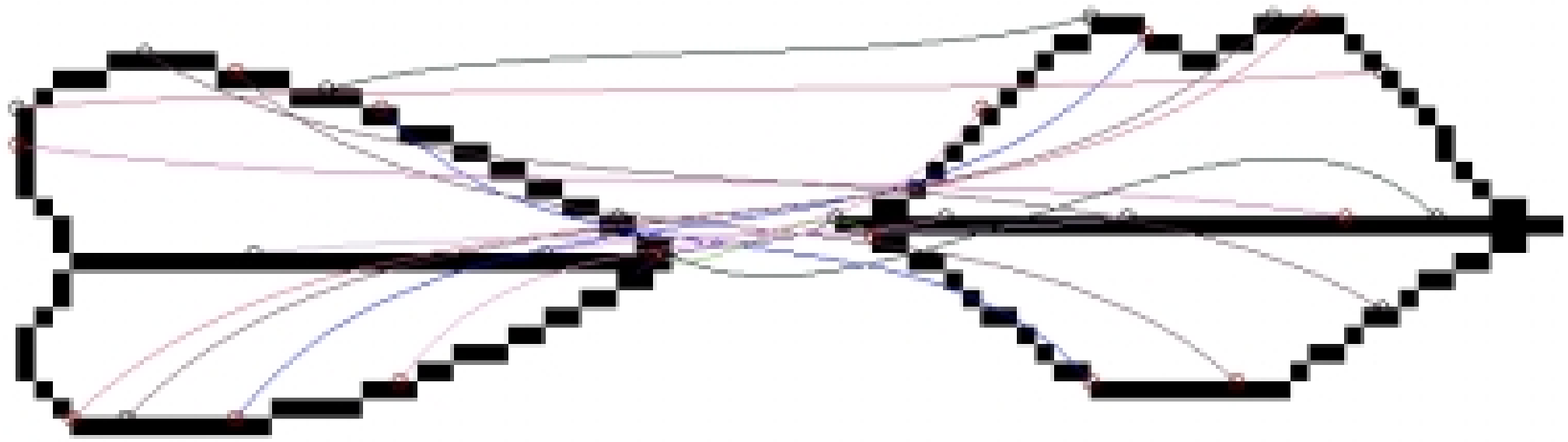

*Figure 33(Before Optimization)*

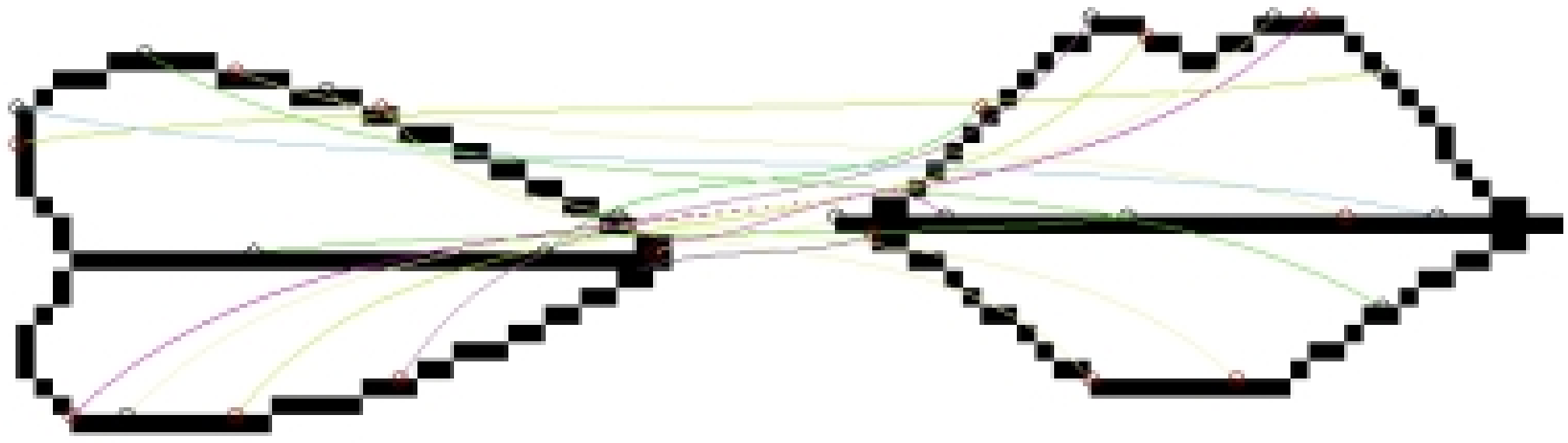

*Figure 34(After Optimization)*

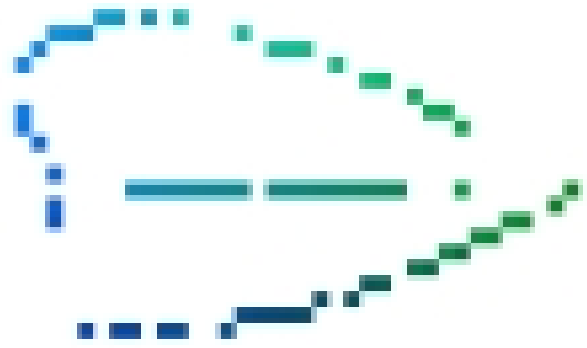

*Figure 35*

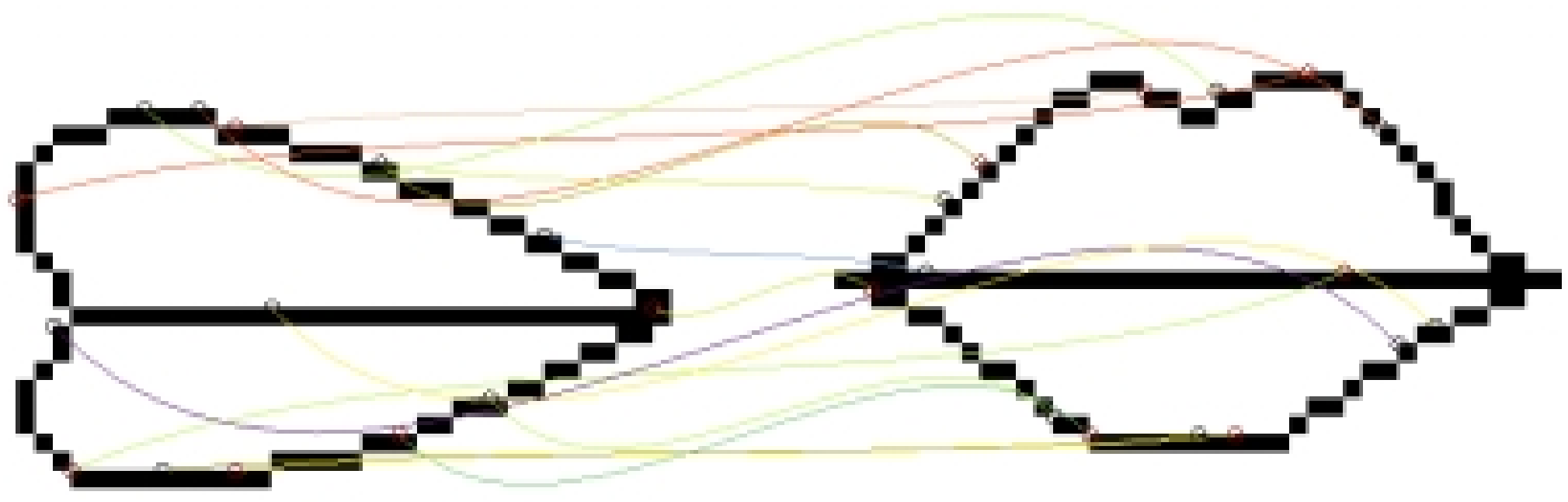

*Figure 36(Before Optimization)*

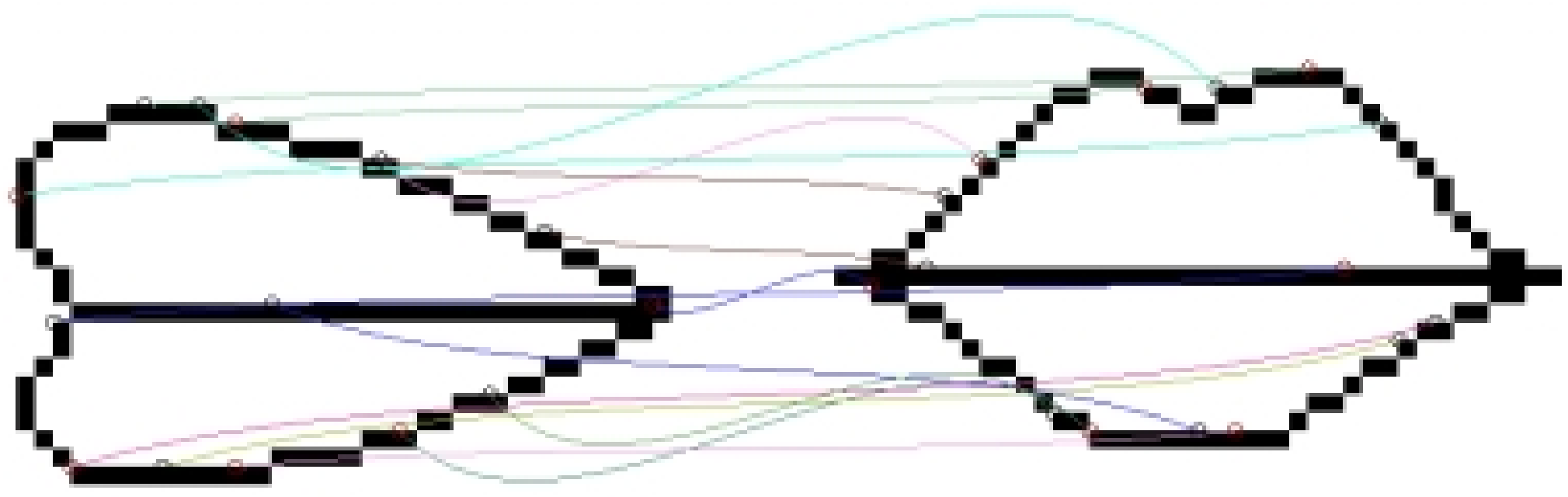

*Figure 37(After Optimization)*

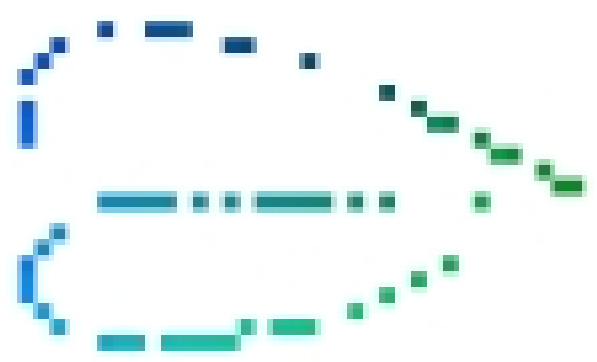

*Figure 38*

In all the presented simulations, an automatic procedure was used to select source and target image's points. By a comparison between Figures 35, and 38, and from the previous examples also, we see clearly, that the numerical simulations confirm to a great degree our point of view.

## IV. Conclusion, discussion, and development for the Future.

Explaining the philosophical details, which led to the presented model, is beyond the space provided by this paper. The essential in the presented model, is the exploration of two important aspects, the “Object” aspect, and the “Analysis” aspect. Analysis, as we presented in the previous sections, is nothing but a recursive procedure working on the sub-parts of the object in a way consistent and compatible with its application on the main root part. Numerical simulations do confirm to a great degree our point of view. At the same time, the way in which we proceeded approaching our model is just an opening, and more development is needed in the future, especially concerning the following points:

- Numerical Optimization: the space of solutions is of extreme possibilities. Beside more powerful calculation machines, a more adaptive, more intelligent algorithm for the space exploration can be developed in the future. Probabilistic models can be built-on the presented measure, to economize the needed number of calculation.
- Many-to-one and One-to-many Model: For simplicity, we limited our exploration, only to a one-to-one model (2), while model (1) is more realistic and accurate. At the same time model (1) is very costly in calculus volume.
- High Level Object Model: the optimal solution between two objects is not necessarily the final solution, i.e. when these two objects are composed of resembling sub-objects in totally misplaced locations. A global view of the most significant minimums must be always preserved, and a more Object-Oriented, search strategies must be developed, according to the intended application. As an example, an automated environment perception system must be capable of finding the Object isolation borders, beyond which, object existence is a contextual fact that is not relevant to object definition. In model (2) vocabulary, this corresponds, to putting zeros in Auto-Distance matrices in the places connecting the “Object” point to the contextual points.